\documentclass[twocolumn,prl,superscriptaddress,showpacs,amsmath]{revtex4}

\usepackage{graphicx}

\begin{document}

\title
{Cavity-enhanced light scattering in optical lattices to probe
atomic quantum statistics}
\date{\today}

\author{Igor B. Mekhov}
\email{Igor.Mekhov@uibk.ac.at} \affiliation{Institut f\"ur
Theoretische Physik, Universit\"at Innsbruck, Innsbruck, Austria}
\affiliation{St. Petersburg State University, V. A. Fock Institute
of Physics, St. Petersburg, Russia}
\author{Christoph Maschler}
\author{Helmut Ritsch}
\affiliation{Institut f\"ur Theoretische Physik, Universit\"at
Innsbruck, Innsbruck, Austria}

\begin{abstract}
Different quantum states of atoms in optical lattices can be
nondestructively monitored by off-resonant collective light
scattering into a cavity. Angle resolved measurements of photon
number and variance give information about atom-number fluctuations
and pair correlations without single-site access. Observation at
angles of diffraction minima provides information on quantum
fluctuations insensitive to classical noise. For transverse probing,
no photon is scattered into a cavity from a Mott insulator phase,
while the photon number is proportional to the atom number for a
superfluid.
\end{abstract}


\maketitle

Studies of ultracold atoms in optical lattices link various
disciplines. Fundamental quantum many-body theories, formulated
initially for condensed matter, can be tested in better controllable
atomic systems~\cite{JakschBloch}, e.g., strongly correlated phases,
quantum simulators. Such studies influence different
areas~\cite{JakschBloch}: quantum information processing, ultracold
collisions, exotic molecules, etc.

While mean-field approaches describe only the average atomic
density, the main goal is to study quantum properties of these
gases. They are most prominent in lattices, where one has phase
transitions between states of similar density but radically
different quantum fluctuations.

Standard methods to measure quantum properties are based on
matter-wave interference of atoms released from a
trap~\cite{blochnewlukin} destroying the system. ``Bragg
spectroscopy'' using stimulated matter-wave scattering by laser
pulses proved successful~\cite{stenger,stoferle} but destructive.
Alternative less destructive methods observing scattered light were
proposed mainly for homogeneous Bose-Einstein condensates (BEC)
\cite{BEC,BECx,BEC2,moore}, but not yet implemented.

Here we show that specifically for periodic lattices, light
scattering can help to overcome experimental difficulties. In
contrast to homogeneous BECs, scattering from a lattice allows to
determine local and nonlocal correlations without single-atom
optical access using the suppression of strong classical scattering
at Bragg minima and monitoring much richer angular distributions.
This looks extremely useful for studying phase transitions between,
e.g., Mott insulator (MI) and superfluid (SF) states, without
destruction, since various quantum phases show even qualitatively
distinct scattering.

Joining two fields, cavity quantum electrodynamics (QED) and
ultracold gases, will enable new investigations of both light and
matter at ultimate quantum levels, which only recently became
experimentally possible~\cite{EsslingerVuletic}.

Our model is based on nonresonant interaction, not relying on a
particle level structure. Thus it also applies to molecular physics,
where new quantum phases were obtained~\cite{RempeGrimm}. It can be
also applied for semiconductors~\cite{Koch}, as, e.g., were used for
BEC of exciton-polaritons~\cite{LeSiDang}.

\begin{figure}
\scalebox{0.6}[0.6]{\includegraphics{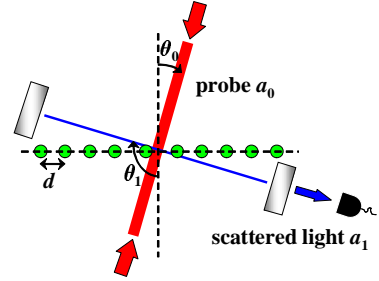}}
\caption{\label{fig1}(Color online) Setup. A lattice is illuminated
by a probe at the angle $\theta_0$ which is scattered into a cavity
at $\theta_1$.}
\end{figure}

{\bf Model}. We consider $N$ two-level atoms in an optical lattice
with $M$ sites. A region of $K\le M$ sites is illuminated by probe
light which is scattered into another mode (cf. Fig.~\ref{fig1}).
Although, each mode could be a freely propagating field, we will
consider cavity modes whose geometries (i.e. axis directions or
wavelengths) can be varied. A related manybody Hamiltonian is given
by

\begin{eqnarray}\label{1}
H=\sum_{l=0,1}{\hbar\omega_l a^\dag_l a_l} + \int{d^3{\bf
r}\Psi^\dag({\bf r})H_{a1}\Psi({\bf r})}, \\
H_{a1}=\frac{{\bf p}^2}{2m_a}+V_{\text {cl}}({\bf r})+\hbar
g^2_0\sum_{l,m=0,1}{\frac{u_l^*({\bf r})u_m({\bf r}) a^\dag_l
a_m}{\Delta_{ma}}},\nonumber
\end{eqnarray}
where $a_0$ ($a_1$) are the annihilation operators of the probe
(scattered) light with the frequencies $\omega_{0,1}$, wave vectors
${\bf k}_{0,1}$, and mode functions $u_{0,1}({\bf r})$; $\Psi({\bf
r})$ is the atom-field operator. In the effective single-atom
Hamiltonian $H_{a1}$, ${\bf p}$ and ${\bf r}$ are the momentum and
position operators of an atom of mass $m_a$ trapped in a classical
potential $V_{\text {cl}}({\bf r})$; $g_0$ is the atom--light
coupling constant. We consider the field-atom detunings $\Delta_{la}
= \omega_l -\omega_a$ larger than the spontaneous emission rate and
Rabi frequencies. Thus, in $H_{a1}$ the adiabatic elimination of the
upper state was used.

Assuming weak fields $a_{0,1}$, we expand $\Psi({\bf r})$ in
Eq.~(\ref{1}) using localized Wannier functions corresponding to
$V_{\text {cl}}({\bf r})$ and keep only the lowest vibrational state
at each site: $\Psi({\bf r})=\sum_{i=1}^{M}{b_i w({\bf r}-{\bf
r}_i)}$, where $b_i$ is the atom annihilation operator at site with
coordinate ${\bf r}_i$. Substituting this in Eq.~(\ref{1}), one can
get a generalized Bose-Hubbard Hamiltonian \cite{JakschBloch}
including light scattering. However, in contrast to our previous
work \cite{maschler} and ``Bragg spectroscopy'' \cite{stoferle}, we
do not consider lattice excitations here and focus on scattering
from atoms in a prescribed state.

Neglecting atomic tunneling, the Hamiltonian reads:

\begin{eqnarray}
H=\sum_{l=0,1}{\hbar\omega_l a^\dag_l a_l}+ \hbar g^2_0
\sum_{l,m=0,1}{\frac{a^\dag_l
a_m}{\Delta_{ma}}}\left(\sum_{i=1}^K{J_{i,i}^{lm}\hat{n}_i}\right),
\nonumber
\end{eqnarray}
where $\hat{n}_i=b_i^\dag b_i$. We define the operator of the atom
number at illuminated sites as $\hat{N}_K=\sum_{i=1}^K{\hat{n}_i}$.
For a deep lattice the coefficients $J_{i,i}^{lm}=\int{d{\bf
r}}w^2({\bf r}-{\bf r}_i) u_l^*({\bf r})u_m({\bf r})$ reduce to
$J_{i,i}^{lm}=u_l^*({\bf r}_i)u_m({\bf r}_i)$ neglecting atom
spreading, which can be studied even by classical
scattering~\cite{Slama}.

The Heisenberg equation for the scattered light in the frame
rotating with $\omega_0$ ($\Delta_{01}=\omega_0-\omega_1$) thus reads:

\begin{eqnarray}\label{2}
\dot{a}_1= i\left[\Delta_{01}
-\frac{g_0^2}{\Delta_{1a}}\sum_{i=1}^K{|u_1({\bf
r}_i)|^2\hat{n}_i}\right]a_1  \nonumber\\
-i\frac{g_0^2 a_0}{\Delta_{0a}}\sum_{i=1}^K{u_1^*({\bf r}_i)u_0({\bf
r}_i)\hat{n}_i}-\kappa a_1,
\end{eqnarray}
where $\kappa$ is the cavity decay rate and $a_0$ will be assumed a
classical field given by a c-number constant.

{\bf Light properties}. Though the dispersion shift of a cavity mode
is sensitive to atom statistics through $\hat{n}_i$, we assume it is
much smaller than $\kappa$ or $\Delta_{01}$. A stationary solution
of Eq.~(\ref{2}) for $a_1$ and photon number then reads

\begin{eqnarray}\label{3}
a_1=C\hat{D}, \quad n_{\text ph}= a^\dag_1a_1=|C|^2\hat{D}^*\hat{D},
\quad \hat{D}=\sum_{i=1}^K{A_i\hat{n}_i},
\end{eqnarray}
with $C\equiv ig_0^2 a_0 /[\Delta_{0a} (i\Delta_{01}-\kappa)]$ and
the coefficients $A_i(\theta_0,\theta_1)\equiv u_1^*({\bf
r}_i)u_0({\bf r}_i)$. This expression of the light operators through
the atomic ones is a central result here.

For a 1D lattice of period $d$ and atoms trapped at $x_m=md$
($m=1,2,...,M$) the mode functions are $u_{0,1}({\bf r}_m)=\exp
(imk_{0,1x}d)$ for traveling and $u_{0,1}({\bf r}_m)=\cos
(mk_{0,1x}d)$ for standing waves with $k_{0,1x}=|{\bf
k}_{0,1}|\sin\theta_{0,1}$ (cf. Fig.~\ref{fig1}). For the atomic
quantum state we use the assumptions: (i) the mean atom number at
all sites is $\langle\hat{n}_i\rangle = n=N/M$
($\langle\hat{N}_K\rangle=N_K\equiv nK$) and (ii) the pair
correlations $\langle\hat{n}_i\hat{n}_j\rangle$ are identical for
any $i\ne j$, which is valid for a deep lattice, and will be denoted
as $\langle\hat{n}_a\hat{n}_b\rangle$ (with $a\ne b$).

Thus, $\langle a_1 \rangle \sim \langle \hat{D} \rangle=
\sum_{i=1}^K{A_i\langle\hat{n}_i}\rangle= nA$ showing that the field
amplitude only depends on the mean density and exhibits the angular
distribution of classical diffraction $A(\theta_0,\theta_1)\equiv
\sum_{i=1}^K{A_i(\theta_0,\theta_1)}$ with diffraction maxima and
minima. The central point now is that the photon number (\ref{3}) is
not just the amplitude squared, but we get

\begin{subequations}\label{4}
\begin{eqnarray}
\langle \hat{D}^*\hat{D} \rangle =\langle \hat{n}_a\hat{n}_b\rangle
|A|^2+(\langle\hat{n}^2\rangle -
\langle \hat{n}_a\hat{n}_b\rangle)\sum_{i=1}^K{|A_i|^2}, \label{4a}\\
R(\theta_0, \theta_1)\equiv \langle\hat{D}^*\hat{D} \rangle - |\langle \hat{D}
\rangle|^2 = \nonumber\\
=\langle \delta\hat{n}_a\delta\hat{n}_b\rangle
|A|^2+(\langle\delta\hat{n}^2\rangle - \langle
\delta\hat{n}_a\delta\hat{n}_b\rangle)\sum_{i=1}^K{|A_i|^2},\label{4b}
\end{eqnarray}
\end{subequations}
where $\delta\hat{n}_i\equiv\hat{n}_i - n$ giving
$\langle\delta\hat{n}_a\delta\hat{n}_b\rangle=\langle
\hat{n}_a\hat{n}_b\rangle-n^2$, and $\langle\delta\hat{n}^2\rangle$
equal to the variance $(\Delta
n_i)^2=\langle\hat{n}_i^2\rangle-n^2$. Thus, the intensity is
sensitive to atomic quantum statistics via the density-density
correlations $\langle\hat{n}_i\hat{n}_j\rangle$ different for
particular states. Besides the classical angle dependence $|A|^2$,
the second term in Eq.~(\ref{4a}) reflects fluctuations and has a
completely different dependence. Particularly in a lattice,
scattering is sensitive not only to the periodic density, but also
to periodic fluctuations, leading to the observable difference
between states with and without nonlocal correlations. Analysis of
quadrature variances gives results similar to analysis of the noise
quantity $R$.

For two traveling waves, Eq.~(\ref{4a}) gives the structure factor
considered in Ref.~\cite{BEC2} on homogeneous BECs. We show that a
more general form including standing waves gives new measurable
quantities beyond structure factor.

The intensity fluctuations of the scattered light depend on the
fourth moments of the atomic number operators and four-point density
correlations $\langle \hat{n}_i\hat{n}_j\hat{n}_k\hat{n}_l\rangle$.
For example, the photon-number variance is given by $(\Delta
n_\text{ph})^2=\langle n_\text{ph}^2\rangle - \langle
n_\text{ph}\rangle^2=|C|^4(\langle |\hat{D}|^4\rangle -\langle
|\hat{D}|^2\rangle^2)+|C|^2\langle |\hat{D}|^2\rangle$.

To discuss examples of different scattering we summarize statistical
properties of typical states in Table~\ref{table1}. For light
scattering, the most classical state corresponding to pointlike
atoms is MI. Here the atom number at each site $\hat{n}_i$ does not
fluctuate and we have no pair correlations. Hence we see from
Eq.~(\ref{4a}) that the zeros of classical diffraction
[$A(\theta_0,\theta_1)=0$] are zeros of light intensity.

\begin{table}
\begin{tabular}{|l|l|l|l|}\hline
& {\bf MI} & {\bf SF} & {\bf Coherent}\\ \hline \hline
$|\Psi\rangle$ & $\displaystyle\prod_{i=1}^M |n_i\rangle_i$ &
$\displaystyle\frac{1}{\sqrt{M^N N!}}(\sum_{i=1}^M
b_i^\dag)^N|0\rangle$ & $\displaystyle
e^{-\frac{N}{2}}\prod_{i=1}^M{e^{\sqrt{\frac{N}{M}}b_i^\dag}}
|0\rangle_i$
\\\hline $\langle\hat{n}_i^2\rangle$ &  $n^2$   & $n^2(1-1/N)+n$ &
$n^2+n$
\\\hline
$\left(\Delta n_i\right)^2$ &  0 & $n(1-1/M)$ & $n$
\\\hline
$\langle \hat{N}_K^2\rangle$ & $N_K^2$ & $N_K^2(1-1/N)+N_K$ &
$N_K^2+N_K$ \\\hline $(\Delta N_K)^2$ &  0 & $N_K(1-K/M)$ & $N_K$
\\\hline
$\langle\hat{n}_a\hat{n}_b\rangle$ & $n^2$ & $n^2(1-1/N)$ & $n^2$
\\\hline
$\langle\delta \hat{n}_a\delta \hat{n}_b\rangle$ & 0 & $-N/M^2$ & 0
\\\hline
\end{tabular}
\caption{\label{table1}Statistical quantities of typical atomic
states.}
\end{table}

This is different for a SF where each atom is delocalized over all
sites leading to number fluctuations at a given site and at $K<M$
sites; the atoms at different sites are anticorrelated. At a
classical diffraction zero we still find a photon number
proportional to the atom number $N$.

A coherent state approximates a SF but without pair correlations. In
the limit $N,M\rightarrow \infty$, it well describes scattering from
a small region ($K\ll M$) of a partially illuminated superfluid
(SF$_K$). However, we proved that even in this limit it fails to
describe scattering at angles of Bragg maxima from a large lattice
region ($K\sim M$).

{\bf Example.} Let us now show the most striking predictions of this
model at the basic example of a probe transverse to the lattice
($\theta_0=0$ cf. Fig.~\ref{fig1}). The scattered light is collected
in a cavity along the lattice ($\theta_1=\pi/2$) with atoms trapped
at the antinodes ($d=\lambda/2$)
\cite{maschler,domokos2002ccablack2005clf}.

The operator $\hat{D}=\sum_{k=1}^K(-1)^{k+1}\hat{n}_k$ (\ref{3})
here gives almost zero average field amplitude independently on the
atomic state. This reflects the opposite phase of light scattered
from atoms separated by $\lambda/2$ (diffraction minimum). However,
the cavity photon-number is proportional to
$\langle\hat{D}^*\hat{D}\rangle =(\langle\hat{n}^2\rangle - \langle
\hat{n}_a\hat{n}_b\rangle)K$ [cf. Eq.~(\ref{4a})], which is
determined by statistics of a particular state. Thus, atoms in a MI
state scatter no photons, while a SF scatters number of photons
proportional to the atom number:
\begin{eqnarray}
\quad \langle a_1\rangle_\text{MI}&=&\langle a_1\rangle_\text{SF}=0, \quad \text{but} \nonumber\\
\langle a_1^\dag a_1\rangle_\text{MI}&=&0, \quad \langle a_1^\dag
a_1\rangle_\text{SF}=|C|^2N_K . \nonumber
\end{eqnarray}

Moreover, the photon number fluctuations $(\Delta n_\text{ph})^2$
are also different for various atomic states. In the MI state, the
variance $(\Delta |D|^2)^2_\text{MI}=\langle
|\hat{D}|^4\rangle_\text{MI} -\langle
|\hat{D}|^2\rangle^2_\text{MI}=0$, whereas in SF, there is a strong
noise $(\Delta |D|^2)^2_\text{SF}\approx 2N_K^2$.

Coupled light-matter dynamics in a cavity can lead to a new
self-organized phase \cite{domokos2002ccablack2005clf} with atoms
trapped at every second site ($d=\lambda$), which gives $\hat{D} =
\sum_{k=1}^K\hat{n}_k=\hat{N}_K$ (\ref{3}). If this state is a MI
with $d=\lambda$, the number of photons $\langle a_1^\dag
a_1\rangle_\text{Self-org}=|C|^2 N_K^2$ is proportional to the atom
number squared and has a superradiant character.

\begin{figure}
\scalebox{1.0}[1.0]{\includegraphics{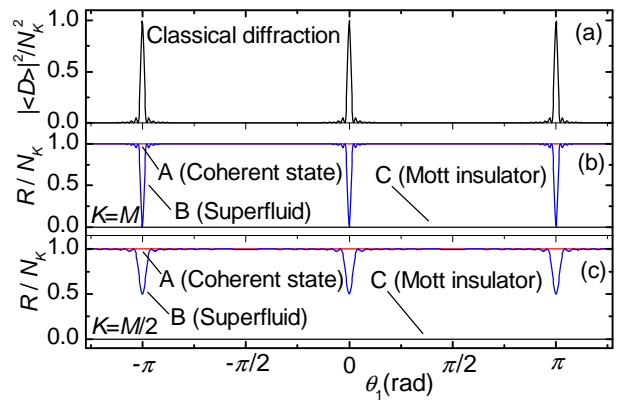}}
\caption{\label{fig2}(Color online) Intensity angular distributions
for two traveling waves. (a) Intensity of classical diffraction; (b)
noise quantity $R$ (\ref{5}) for coherent atomic state (constant 1,
line A), SF with all sites illuminated $K=M$ (curve B), and MI state
(constant 0, line C); (c) the same as in (b) but for partially
illuminated SF with $K=M/2$. $N=M=30$, $\theta_0=0$.}
\end{figure}

\begin{figure}
\scalebox{1.0}[1.0]{\includegraphics{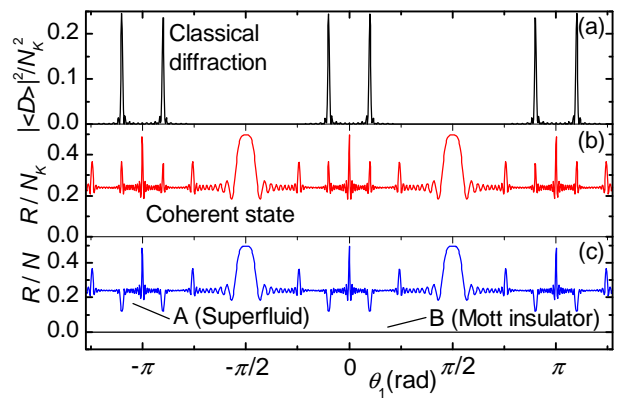}}
\caption{\label{fig3}(Color online) Intensity angular distributions
for two standing-wave modes. (a) Intensity of classical diffraction;
(b) noise quantity for coherent state; and (c) for SF (curve A) and
MI (constant 0, line B). $N=M=K=30$, $\theta_0=0.1\pi$.}
\end{figure}

{\bf Angular distributions.} We will quantitatively discuss angular
intensity distributions for scattering between two traveling waves,
where Eq.~(\ref{4b}) reduces to

\begin{eqnarray}\label{5}
R = \langle \delta\hat{n}_a\delta\hat{n}_b\rangle
\frac{\sin^2{(K\alpha_-/2})}{\sin^2{(\alpha_-/2})}
+(\langle\delta\hat{n}^2\rangle - \langle
\delta\hat{n}_a\delta\hat{n}_b\rangle)K.
\end{eqnarray}

While $|A|^2$ in the first term reproduces classical diffraction
with $\alpha_-=k_{0x}d\sin\theta_0 - k_{1x}d\sin\theta_1$, the
second term in Eq.~(\ref{4b}) is simply isotropic. Thus, the noise
quantity is zero for MI, $R_\text{MI}=0$, nonzero but isotropic for
the coherent state, $R_\text{Coh}=N_K$, and angle dependent for a
SF. In a SF, even small pair correlations $\langle
\delta\hat{n}_a\delta\hat{n}_b\rangle=-N/M^2$ give a large
contribution near diffraction maxima ($\alpha_-=2\pi l, l=0,1,..$),
where the geometric factor is $K^2$, invalidating the coherent-state
approximation.

Figure~\ref{fig2} displays those angular distributions. Classical
diffraction $|\langle D\rangle|^2$ with the only possible zero-order
maxima at $\theta_1=0, \pi$ ($d=\lambda_{0,1}/2$, $\theta_0=0$) is
shown in Fig.\ref{fig2}(a). $R$ for the coherent and SF$_K$ states
are plotted in Figs.~\ref{fig2}(b), (c). For MI, $R=0$. In SF, there
is a noise suppression at maxima, which is total for all sites
illuminated, $K=M$, and partial for $K=M/2$.

In a maximum, $\hat{D}$ (\ref{3}), is reduced to $\hat{N}_K$. Thus,
the field amplitude is determined by $N_K=nK$, the intensity depends
on $\langle\hat{D}^*\hat{D} \rangle=\langle \hat{N}_K^2\rangle$,
while $R=(\Delta N_K)^2$ gives the atom-number variance at $K$
sites, which reflects the total and partial noise suppression in
Figs.~\ref{fig2}(b) and \ref{fig2}(c), since $\langle N_K\rangle$
fluctuates for $K<M$. In diffraction minima, the field is zero, but
the intensity is proportional to $\langle\hat{n}^2\rangle - \langle
\hat{n}_a\hat{n}_b\rangle$. Under scattering of spatially incoherent
light, the intensity is isotropic and proportional to $\langle
\hat{n}^2\rangle$.

So, in optical experiments, varying the geometry, the global
statistics of $K$ sites, local single-site statistics, and pair
correlations can be obtained even without a single-site access.
Thus, light scattering gives a way to distinguish between atomic
states. As shown by Eq.~(\ref{5}) and Fig.~\ref{fig2}, MI and SF$_M$
states are different in diffraction minima and in incoherent light.
They are indistinguishable in maxima. SF$_M$ and coherent states
differ in maxima only. The MI and coherent state are different at
any angles.

The noise quantity or photon statistics are different in orders of
$N_K$ for various states. Nevertheless, for large $N_K$, there could
be practical problems to subtract large values in a maximum. In
Refs.~\cite{BECx}, this even led to a conclusion about the state
indistinguishability by intensity measurements. In contrast to
homogeneous BECs, in lattices, this problem has a natural solution:
measurements outside maxima are free of the strong classical-like
part of scattering and thus directly reflect fluctuations.

A classical analogy of different light scattering consists in
different density fluctuations. A quantum treatment gives a deeper
insight. Superfluid state is a superposition of all possible
multisite Fock states giving distributions of $N$ atoms at $M$
sites. Various Fock states become entangled to scattered light of
different phases and amplitudes. In contrast to a classical case
(and MI with the only multisite Fock state), light fields entangled
to various distributions do not interfere with each other due to the
orthogonality of the Fock states. This reflects the which-way
information and explains the zero amplitude but nonzero photon
number in a diffraction minimum.

If at least one of the modes is a standing wave, the angle
dependences become much richer. Besides new classical maxima given
by $\alpha_{\pm}=k_{0x}d\sin\theta_0 \pm k_{1x}d\sin\theta_1$, the
second, ``noise,'' term in Eqs.~(\ref{4a}) and (\ref{4b}) is also
not isotropic. It includes a sum of the geometric coefficients
squared, which is equivalent to effective doubling of the lattice
period (or light frequency doubling) leading to new features at
angles, where classical diffraction predicts zero. In
Fig.~\ref{fig3}, a case of two standing waves is shown. Due to the
effective period doubling (given by
$2\alpha_{0,1}=2k_{0,1x}d\sin\theta_{0,1}$ and $2\alpha_{\pm}$), new
features at the angles of, e.g., effective first-order maxima
appear, though classically only the zero-order maxima are still
possible.

The angle dependence of the photon number variance $(\Delta
n_\text{ph})^2$ determined by $(\Delta |D|^2)^2$ shows anisotropic
features due to ``period doubling'' even for two traveling waves.
For the coherent state, the light at a maximum displays strong noise
[$(\Delta |D|^2)^2=4N_K^3+6N_K^2+N_k$ because $\langle
|\hat{D}|^4\rangle=N_K^4+6N_K^3+7N_K^2+N_k$ and $\langle
|\hat{D}|^2\rangle=N_K^2+N_K$], stronger than the isotropic
component ($N_K^2$ in highest order of $N_K$) and new features at
$\theta_1=\pm \pi/2$ (for $\theta_0=0$, $2N_K^2$ in highest order of
$N_K$). In SF$_M$, the noise at maxima can be suppressed, while at
other angles it is nearly equal to that of the coherent state. In
MI, $(\Delta |D|^2)^2=0$. Distinguishing between atomic states by
light statistics is similar to that by the intensity.

In summary, we have shown that atomic quantum states can be
nondestructively monitored by measuring scattered light. In contrast
to homogeneous BECs, scattering from lattices exhibits advantageous
properties: suppression of the classical scattering in Bragg minima,
access to local and nonlocal correlations, angular distributions
richer than classical diffraction. Also, other optical phenomena and
quantities depending nonlinearly on the atom number operators will
reflect quantum atom statistics \cite{Meystre,ICAP,arxiv}, e.g., the
dispersion of a medium will provide a spectral method of quantum
state characterization~\cite{WeNaturePhys}. Exploiting quantum
properties of light should be applicable to other Bragg spectroscopy
setups.

Support by: Austrian Science Fund (P17709, S1512).


\end{document}